# Optical excitation and detection of neuronal activity

*Chenfei Hu[1,6], Richard Sam[2,6], Mingguang Shan[6,7], Viorel Nastasa[6,8,#], Minqi Wang[2,6], Taewoo Kim[1,6,†], Martha Gillette[3,4,5,6], Parijat Sengupta[3,5,6], and Gabriel Popescu[1,3,6,*]*

1) Department of Electrical and Computer Engineering,
2) School of Molecular and Cellular Biology,
3) Department of Bioengineering,
4) Department of Cell and Developmental Biology,
5) Neuroscience Program,
6) Beckman Institute for Advanced Science and Technology,
University of Illinois at Urbana-Champaign, Urbana, Illinois 61801, USA
7) College of Information and Communication Engineering,
Harbin Engineering University, Harbin, Heilongjiang 150001, China
8) National Institute for Laser Plasma and Radiation Physics, Magurele, Ilfov 077125, Bucharest, Romania

†Currently with Caltech Optical Imaging Laboratory, California Institute of Technology, Pasadena, CA, 91125, USA.

# Currently at Extreme Light Infrastructure - Nuclear Physics/ Horia Hulubei National Institute for R&D in Physics and Nuclear Engineering, 30 Reactorului Street, 077125 Magurele, Ilfov, Romania

*Corresponding authors: gpopescu@illinois.edu and parijat@illinois.edu

## ABSTRACT

Optogenetics has emerged as an exciting tool for manipulating neural activity, which in turn, can modulate behavior in live organisms. However, detecting the response to the optical stimulation requires electrophysiology with physical contact or fluorescent imaging at target locations, which is often limited by photobleaching and phototoxicity. In this paper, we show that phase imaging can report the intracellular transport induced by optogenetic stimulation. We developed a multimodal instrument that can both stimulate cells with subcellular spatial resolution and detect optical pathlength changes with nanometer scale sensitivity. We found that optical pathlength fluctuations following stimulation are consistent with active organelle transport. Furthermore, the results indicate a broadening in the transport velocity distribution, which is significantly higher in stimulated cells compared to optogenetically inactive cells. It is likely that this label-free, contactless measurement of optogenetic response will provide an enabling approach to neuroscience.

**Introduction**

Optogenetics is a transforming tool in the field of cellular biology and especially neuroscience. By genetic engineering, a cell can express light-sensitive proteins, and then its activity can be initiated or inhibited using light. For example, delivering light stimulation on *opsin* family proteins causes changes in the cation level, which can trigger or prevent action potentials in neurons [1]. Unlike electrical stimulation, light stimulation manipulates specific cells that are expressing the targeted opsin, thereby making it possible to investigate the role of a subpopulation of neurons in a neural circuit [2]. Over the past decade, accompanied by advances in virus targeting methods (*i.e.*, adeno-associated virus), as well as novel light delivery mechanisms, the optogenetics toolbox has been growing and gradually becoming a standard method for studying neural functions at both the cellular and behavioral level [2-4]. It is noteworthy that use of optogenetics nowadays extend far beyond neurobiology. Light-sensitive proteins are now being used to control gene expression, cell movement and modification of structure of the cell from the outside. For *in vitro* experiments, light stimulation can be achieved by using the excitation light source in fluorescent microscopes with appropriate wavelength and intensity. By coupling additional components to the light port, patterned illumination delivers arbitrary patterns on the sample, which enables fast, multi-site and high-resolution stimulation. Patterned illumination mostly involves using paired galvo mirrors, Digital micromirror devices (DMD) or Spatial light modulator (SLM) working under one-photon or two-photon excitation strategies [5-8].

Currently, electrophysiological methods are considered to provide the highest fidelity readout of neural activity, which is essentially achieved by attaching physical electrodes to the sample [9]. Though these approaches offer high sensitivity, they require physical contact and cell impaling, while the throughput is low. Recent developments in micro-electrode arrays allows for a simultaneous recording of up to a few hundred neurons [2]. Unfortunately, the low spatial resolution, lack of control over the excitation, and photoelectric effects induced in the electrode by the process of stimulation are inevitable limitations [10]. Optical imaging is a potential solution for circumventing these limitations of electrophysiology. This is typically realized by introducing exogenous fluorescent labels or sensors that change their properties when cells are activated [11, 12]. However, the process requires tedious sample preparation and these fluorescent labels are often toxic to cells [13].

Quantitative phase imaging (QPI) [14] has emerged as a valuable tool for live cell imaging, especially because it is label-free and nondestructive. QPI relies on the principle of interference, whereby an image field is overlaid with a reference field. As a result, even the most transparent objects, such as unlabeled live cells, can be imaged with high contrast and sensitivity using the phase information of the field. Because the phase of the image field is measured quantitatively, it can report on both the thickness and dry mass density of the specimen. A number of methods have been proposed, especially over the past 1-2 decades, to optimize the following properties of phase imaging: spatial and temporal resolution, spatial and temporal sensitivity [15-20]. With the recent advances, QPI has become a significant method for studying live cells, such as red blood cell dynamics [21-24], cell growth [25, 26], cell dynamics [27-30] and cell tomography [31-33]. More recently, due its non-invasive and high sensitivity to sub-nanometer changes in optical pathlength, QPI also found applications in neuroscience, and enabled multi-scale structure and dynamics studies of neurons from a single cell [34, 35] to neuronal network[36, 37], to brain tissue levels[38, 39].

In this paper, we report a novel all-optical method that combined patterned light stimulation of optogenetic cells with functional phase imaging, without labels or physical contact. We also present a new instrument that is capable of exciting optogenetically transformed cells with light and detecting the cell response using interferometry. The overall optical pathlength sensitivity for this instrument is 1.1 nm, which enabled measurement of very small changes in dry mass density. For our experiments, we used PC12 cells derived from rat pheochromocytoma [40], a cell line that is used extensively as a model system for neurons. These cells can be differentiated into a cholinergic neuronal phenotype by nerve growth factor (NGF) treatment [40, 41]. In these cells, NGF induces biochemical, electrophysiological and morphological (neurite outgrowth) changes that are similar to many features characteristic of differentiated sympathetic neurons [42, 43]. Additionally, we demonstrate that the optical pathlength signals measured from PC12 cells using QPI actually reports on enhanced cellular transport that is associated with neuronal cell activation.

**Results**

The experimental setup is shown in Fig. 1. We use a commercial inverted microscope (Axio Observer Z1, Zeiss) onto which we attach the excitation and detection modules. In order to achieve single cell level resolution and selectivity in excitation wavelength, we use a high power commercial projector (Epson Home Cinema 5030UB). The epi-fluorescence excitation source of the commercial microscope is replaced with the commercial projector, which forms an image of the projected pattern on the sample plane via the collector lens (Lens 3) and brightfield objective lens (40X, NA=0.75). The diffraction limited image of this excitation pattern is relayed by the microscope to the camera plane, via the reflection optical path. Dividing the dimension of the projected pattern measured on the camera by the number of pixels, the demagnification factor of this lens system is calculated to be ~10, which indicates a single projector pixel has a dimension around 0.8 μm at the sample plane.

Figure 1b illustrates an excitation dot at blue (centered at 450 nm) wavelength, which is projected onto a live cell. Simultaneously, on the transmission path, we built a phase sensitive interferometric system, known as the diffraction phase microscopy (DPM). This system is described in more detail elsewhere [44, 45]. Briefly, at the image plane of the microscope, we place a diffraction grating, which splits the imaging field into multiple diffraction orders. All the diffraction orders are blocked, except for the $0^{th}$ and $1^{st}$ order. These two beams form an off-axis common-path Mach-Zehnder interferometer, in which Lenses 1-2 form a 4f system that images the grating at the camera plane. In order to obtain a reference field for the interferometer, we spatially filter the $0^{th}$ order through a pinhole at the Fourier plane of Lens 1. The $1^{st}$ order is passed without filtering and carries full information (i.e., amplitude and phase) about the image field. The spatial filter is achieved by an amplitude SLM, onto which we write a binary mask of maximum (white) and minimum (black) transmission, as shown in Fig. 1c.

The camera records an interferogram of the form

$$I(x,y) = I_0 + I_1 + 2\sqrt{I_0 I_1} \cos[\varphi(x,y) + \alpha x] \qquad [1]$$

Where $I_0$ and $I_1$ are the intensities of the (filtered) $0^{th}$ and $1^{st}$ order, respectively, $\varphi$ is the phase map of the object and $\alpha$ is the spatial modulation frequency, $\alpha = 2\pi/\Lambda$, with $\Lambda$ the period of the grating. Ensuring proper sampling (see Ref [45] for details), the quantitative phase map is obtained via a Hibert transform. We chose the DPM system for our phase imaging because the sample and reference field propagate through the same path, thus, highly stable, and it also provides high-throughput due to its single shot performance. Moreover, using a LED as a white-light source with low temporal coherence, this system minimizes the speckle noise that can degrade the sensitivity of the phase imaging system [46]. Figure 1d illustrates an interferogram associated with a live neuron. The three spectral bands of the projectors are shown in Fig. 1e. To eliminate the overlap between the three channels, we used a 450nm band-pass and a 610 nm long pass filters, respectively. The blue channel was used for excitation and the red for control.

Figure 2 shows the phase reconstruction procedure and the noise characteristic of the DPM system. An implementation of spatial Hilbert transform takes the input interferogram (Fig. 2a), performs a Fourier transform (Fig. 2b), selects only one side of the Fourier spectrum (red continuous circle in Fig. 2b), shifts this selection to the center of the image (dotted circle in Fig. 2b), and Fourier transform this signal back to the spatial domain, where the argument provides the phase map, $\varphi$ (Fig. 2c). Details of this process is discussed in Ref [45].

In order to assess the stability of our instrument, which in turn, governs the spatiotemporal sensitivity to optical pathlength changes, we recorded a time series of 'no sample' images (inset of Fig. 2d). The image stack was acquired at 5 frames/s lasted for one minute, with a field of view (FOV) of approximately 35 μm × 45 μm. Each frame was reconstructed, and the histogram of the phase values of the stack was then calculated, as shown in Fig. 2d. The standard deviation of the distribution is computed to be 1.13 nm.

Next, we imaged live NGF-differentiated PC-12 cells using our excitation-detection composite system. The schematic of the experiment is depicted in Fig. 3a. Two types of PC-12 cells were used in our experiment – cells expressing channelrhodopsin-2 (ChR2+ cells) and control (ChR2- cells). ChR2+ cells also expressed a red fluorescent protein, TdTomato, which helped identification of optically active cells (Fig. 3b-c). We performed DPM imaging before and after blue light stimulation, as well as after red light stimulation (See *Materials and Methods* for details). It is known that blue light (peak: 460 nm) [47] stimulation of ChR2 opens up the ion channel in ChR2, which allows positive ions like calcium, potassium, sodium and hydrogen to enter the cell from the extracellular medium. The sudden influx of positive ions depolarizes the cell and activates it. Stimulation with red light does not activate these cells.

The DPM time lapse was acquired at 5 frames/s for a duration of 180 seconds. To observe in detail the dynamic changes associated with the optogenetic excitation, we plot the optical pathlength along a segment of a dendrite vs. time (Fig. 3d). The measurements indicated a clear increase under blue light excitation. This behavior was not observed after red light excitation. Note that the pathlength change that follows cell activation is very subtle, of the order of 10-20 nm, which will be undetectable under a conventional, intensity-based microscope.

Using this procedure, we analyzed 9 regions of interest (ROI) of 5 active and 7 inactive neurons, respectively. The results are summarized in Fig. 3e-f, where we plot the magnitude of change in the optical pathlength with respect to the frame at $t$=0. Clearly, we obtain a significant increase in OPL only for ChR2+ cells after excitation with blue light. Additionally, multi-electrode electrophysiology of cells with blue light stimulation shows induction of activity (Fig. 3g-h).

Cell depolarization can trigger the transport of various cellular organelles, including vesicles, mitochondria and peroxisomes [48-50]. In order to quantify intracellular mass transport, we used dispersion-relation phase spectroscopy (DPS) [27, 37, 51]. This approach allows us to analyze time-resolved phase maps and extract information about the nature of mass transport, namely, diffusive or deterministic. From the image stack, we compute a temporal bandwidth, $\Gamma$, at each spatial frequency, $q$ (see *Materials and Methods* for details). Modeling the intracellular transport by a diffusion-advection equation, the dispersion relation satisfies

$$\Gamma(q) = \Delta v q + D q^2 \qquad [2]$$

where $\Delta v$ is the width of the velocity distribution and $D$ is the diffusion coefficient. Thus, by fitting the data with Eq. 2, we can find out both $\Delta v$ and $D$.

We applied DPS to subcellular regions before and after stimulation with both red and blue light. Figure 4a-d illustrates this analysis for one ChR2+ and one ChR2- cell. On the ROIs containing dendrites or axons, the data shows that deterministic transport (*i.e.,* linear $\Gamma$ vs. $q$ curve) is dominant in all cases as expected. However, upon blue stimulation, there is a 35.8% increase in the $\Delta v$ value in this ChR2+ cell, compared with the value under the condition of no stimulation. This increase is indicative of a high probability for faster transport, irrespective of the velocity direction. This result is consistent with those we obtained in Figs 3e-f. Figure 4e summarizes the $\Delta v$ measurements obtained from different types of cells and conditions. In ChR2- cells, $\Delta v$ values obtained with either blue or red light stimulation are rather similar (*p-value* = 0.00433). In contrast, in ChR2+ cells, excitation with blue light is accompanied by a change in $\Delta v$ value of 25± SE4.2% (n = 12) compared to $\Delta v$ value measured with red light excitation. This change in $\Delta v$ is statistically significant as evidenced by a *p-value* of 0.0043. This result indicates that there is an increase in the width of the velocity distribution when cells are stimulated.

**Discussion**

We developed a multimodal instrument for spatially targetted optical stimulation and label-free non-contact measurement of cell activation which is of utmost importance in cell biology. This stimulation module exploits computer projection to achieve subcellular spatial resolution and can work at multiple channels. The detection path consists of a highly sensitive phase imaging system, which provides sensitivity to optical pathlength changes in the order of 1 nm. Using this instrument, we measured nanometer scale changes in optical pathlength in living NGF-differentiated PC12 cells under basal and stimulated conditions.

NGF induced differentiation of PC12 cells creates a neuronal phenotype that has been widely used as a convenient model system for neurons for cell biological studies of neurotrophin action, monoamine biogenesis, protein trafficking and secretory vesicle dynamics [52]. We expressed ChR2 in these cells to render them optically excitable. Light stimulation of ChR2-expressing PC12 cells has been characterized previously in great detail [53] to demonstrate that these cells are excitable at low photon density [53], and stimulation caused an increase in cytosolic calcium levels [54]. The cells contain large (~100nm) dense core vesicles (DCVs) storing monoamines, and relatively smaller (~40nm) vesicles of endosomal origin termed synaptic vesicle-like microvesicles (SLMVs) that contains acetylcholine [40, 55, 56]. Elevation of cytosolic calcium triggers movement of these vesicles, vesicle fusion with the plasma membrane, and secretion from these vesicles [54-59].

We studied both ChR2+ and ChR2- cells under basal and stimulated conditions. As a negative control with light on, stimulation of ChR2+ cells with red (610 -690 nm) light was used. We found a significant increase in cellular dry mass fluctuations, or optical pathlength, for the optically excited group (ChR2+ cells, blue light). Detailed analysis of the physical nature of these fluctuations reveals a dispersion relation that is consistent with deterministic organelle transport initiated by the optical stimulation. The increase in optical path length, or local dry mass, is likely due to enhanced transport of vesicles and other organelle [60]. Furthermore, the value of the change in the velocity distribution standard deviation is significantly higher in the stimulated neurons, suggesting an increase in directed transport. Measuring a dominant deterministic (directed, active) transport is consistent with previous reports on traffic along dendrites of neurons

[61]. Together, these findings indicate that the organelle transport that accompanies the stimulation can be detected without labels or physical contact using quantitative phase imaging.

Note that cells under basal condition also show some degree of activity which is what we expect to observe. Vesicle and other organelle trafficking occurs all the time in a living cell as this is important to maintain cellular function. However, rate of this transport is significantly enhanced when the cell is stimulated.

We argue that the label-free method described here can be used to measure activation of other cells. For example, activation of endocrine cells, where also vesicle transport follows activation. Additionally, it is known that stimulation of neurons increases rate of vesicle trafficking along it axons. This response is very similar to what we observed here with differentiated PC-12 cells. We, and others, have shown that optical stimulation of optogenetically-transformed neurons increases rate of firing and elevates cytosolic calcium. Therefore, we believe the DPM technique described here can be easily extended to primary neurons in culture.

Our current experiments ran at a low acquisition speed, which could not recover spike signals corresponding to calcium influx or even action potentials. However, by simply employing a fast, sensitive camera or a stronger light source, the frame rate of this imaging system can be significantly increased, and this enables the study of subtle and fast cellular activities (*i.e.*, membrane potential changes)[62], which would be our future study of interest. Because our composite instrument can be attached to an existing microscope, we anticipate that our system, together with other instruments in QPI family, will open a new avenue to explore the function of neural networks, which would be broadly benefit the neuroscience community.

**Materials and Methods**

1. *Cell imaging*

Throughout imaging, neurons were set on a heated stage enclosed with an incubator (Zeiss), which maintained an atmosphere of 37 °C and 5% $CO_2$. For one neuron cell, the imaging was repeated 3 times, with each under a different light excitation condition: no stimulation, red light (~650 nm) and blue light (~460 nm), in this order. Fig. 3a shows the imaging process. In these experiments, the stimulation pattern was a disk of 21 μm in diameter shined on the neuron cell body. The stimulation lasted for 3 seconds, and DPM measurement was immediately followed in the next 3 minutes, at a speed of 5 frames/second. The current system provides light power of 0.31 mW/mm$^2$. Considering the long exposure time, the power exerted on the cells should be enough to activate the ion channels [63]. To minimize correlation between different excitation, a time interval around 5-7 minutes was set between two consecutive measurements. The experiment was conducted in a dark room, and cells were only exposed to the light in the process of stimulation, imaging and observing fluorescent signal.

2. *Sample preparation*

PC12 cells were purchased from ATCC, and maintained in ATCC-formulated RPMI-1640 medium supplemented with 10% horse serum and 5% fetal bovine serum under 5% $CO_2$ environment at 37ºC. For imaging, cells were plated on glass-bottomed petri dishes coated with Fibronectin. Cells were nucleofected in suspension with CAG-hChR2-H134R-tdTomato plasmid, a gift from Karel Svoboda (Addgene plasmid # 28017) [64]. NGF (Nerve Growth Factor, 100ng/ml) was added to the growth medium 24 hours after plating to initiate neuronal differentiation. Medium was replenished with new medium every third day. The cells were allowed to differentiate into neurons for at least 7 days before imaging. The methods were carried out in accordance with the relevant guidelines and all experimental protocols were approved by UIUC's Division of Research Safety.

3. *Dispersion-relation Phase Spectroscopy (DPS)*

Dispersion-relation Phase Spectroscopy (DPS) characterizes the nature of mass transport (i.e. deterministic transport and diffusion) without tracking individual particles. Figure 5 summarizes the DPS procedure. Starting from a time series image (Fig. 5a), a subcellular area (Fig. 5b) is manually selected from the whole FOV. Because a phase map, $\varphi$, is essentially proportional to dry mass density, it is assumed to satisfy the diffusion-advection equation, namely[27]

$$D\nabla^2\varphi(\mathbf{r},t) - \mathbf{v}\cdot\nabla\varphi(\mathbf{r},t) - \frac{\partial}{\partial t}\varphi(\mathbf{r},t) = 0 \qquad [3]$$

where $\mathbf{r}$ the spatial coordinates, $D$ the diffusion coefficient, $\mathbf{v}$ the advection velocity. Taking a Fourier transform with respect to $\mathbf{r}$ (Fig. 5c), we obtain the expression in frequency domain

$$\left(-Dq^2 + i\mathbf{q}\cdot\mathbf{v} - \frac{\partial}{\partial t}\right)\varphi(\mathbf{q},t) = 0 \qquad [4]$$

In Eq. 4, we use the same symbols with different argument for a function and its Fourier transform, i.e., $f(r) \leftrightarrow f(q)$, where $f$ is an arbitrary signal and $\leftrightarrow$ indicates the Fourier transform. Following the calculation in Ref. [27], the temporal autocorrelation, $g$, at each spatial frequency can be modeled as

$$g(q,\tau) = e^{i\mathbf{v}_0\cdot\mathbf{q}\tau}e^{-q\Delta v\tau t - Dq^2\tau} \qquad [5]$$

Here, $\mathbf{v}_0$ represents the mean velocity, and it is considered negligible, and a digital registration is also performed on each image stack to minimize shifting between frames. Thus, we obtain the exponentially decay rate at each spatial frequency $\Gamma(q)$ as seen in Eq. 2. And $\Gamma$ is the standard deviation of the spatial power spectrum (Fig.5d). A radial average is performed to compress the 2D map into 1D line profile (Fig. 5e). The $\Delta v$ and $D$ is then extracted by fitting the data to Eq. 2.

### 4. *Micro-electrode array (MEA) electrophysiology*

Multi-electrode electrophysiology was performed using a Multi-Channel Systems broadband amplifier and at 10 kHz sampling frequency. Samples were kept at 37°C during recording using a heating and perfusion system (ALA Scientific) integrated to the MEA amplifier. Electrical activity from differentiated and optogenetically-transformed cells was measured with blue light stimulation. Data analysis was performed using MC_Rack software (Multi-Channel Systems). The signal was filtered using a 200Hz high-pass second order Butterworth filter. After that, spikes were detected using a threshold set at 6x STDEV of the signal. Number of spikes per second was calculated from this spike train data.

**Data availability**

The data that support the findings of this study are available from the corresponding author upon reasonable request.

**Acknowledgement**

We gratefully appreciated the funding support from National Science Foundation (NSF) STC CBET 0939511, NSF BRAIN EAGER DBI 1450962, and IIP-1353368, PS's start-up funding.

**Author Contributions**

M.G., P.S., and G.P. proposed the project. T.K., M.S., and V.N. built the excitation module. C.H. and M.S built the DPM system. M.W. prepared DNA. R.S. prepared cell samples. C.H.

performed quantitative phase imaging and data analysis. G.P. and P.S. supervised the research. C.H, P.S. and G.P. wrote the manuscript with input from all authors.

**Competing financial interest**

G.P. has financial interest in Phi Optics, Inc., a company developing quantitative phase imaging technology for materials and life science applications.

**Figure Captions:**

1. System Schematic. (a) white-light Diffraction Phase Microscopy (wDPM) system combines a projector to achieve spatially-resolved optical stimulation and label-free imaging. (b) Stimulation pattern, a disk of 21 μm in diameter. (c) SLM mask. (d) Raw image of a neuron cell, and the dashed circle indicates the position and size of the stimulation spot. (e) Spectrum of blue, red light stimulation, and fluorescent excitation measured at the sample plane.

2. Procedure of phase reconstruction. (a) a raw image which is an interference between reference and $1^{st}$ order diffraction, the zoom-in image shows the fringes. (b) Fourier transform of the raw image, one of the sidelobes is isolated and then moved to the image center. (c) an optical pathlength (OPL) map is reconstructed after taking inverse Fourier transform, background subtraction and halo removal (unit in nm). (d) Histogram of spatiotemporal noise with a standard deviation of 1.13 nm, the inset represents a time stack images of system noise.

3. (a) Schematic of the imaging process. (b) One fluorescent image of an ChR2 positive cell and (c) its corresponding quantitative phase image with units in OPL. (d) The mass density steadily increased along this active cell dendrite after stimulation with blue light. (e-f) OPL change after different stimulation on both ChR2+ and ChR2- neurons. 9 different ROIs with a size of 30×30 pixels across 5 ChR2 active cells and 7 ChR2 negative cells were selected, respectively. All ROIs were located either on neuron dendrites, axon, or the region of cellbody close to a dendrite. The absolute change of averaged phase with respect to t=0 were plotted for each stimulation condition, with standard errors in light-gray lines. (g)-(h) Multi-electrode electrophysiology signal of cells after blue light stimulation.

4. DPS analysis were performed on one ChR2 positive (a-c) and ChR2 negative(c-d) cells at selected region indicated by a red box for each stimulation condition. (e) Box chart of the transport velocity change compared with the velocity under no light stimulation after different stimulation on ChR2+ and ChR2- (unit in %). A total of 12 ROIs were selected from 5 positive and 7 negative neurons.

5. Procedure of DPS analysis. (a) A region of interest (ROI) in the field of view is selected, and (b) the image volume is digitally aligned to minimize drifts between frames. (c) Taking a Fourier transform on each frame, (d) the temporal bandwidth is then evaluated and obtain map of $\Gamma$. (e) Performing a radial average on (d) and present in a log plot, the advection and diffusion coefficients are extracted from dispersion curve.

**Figures:**

1.

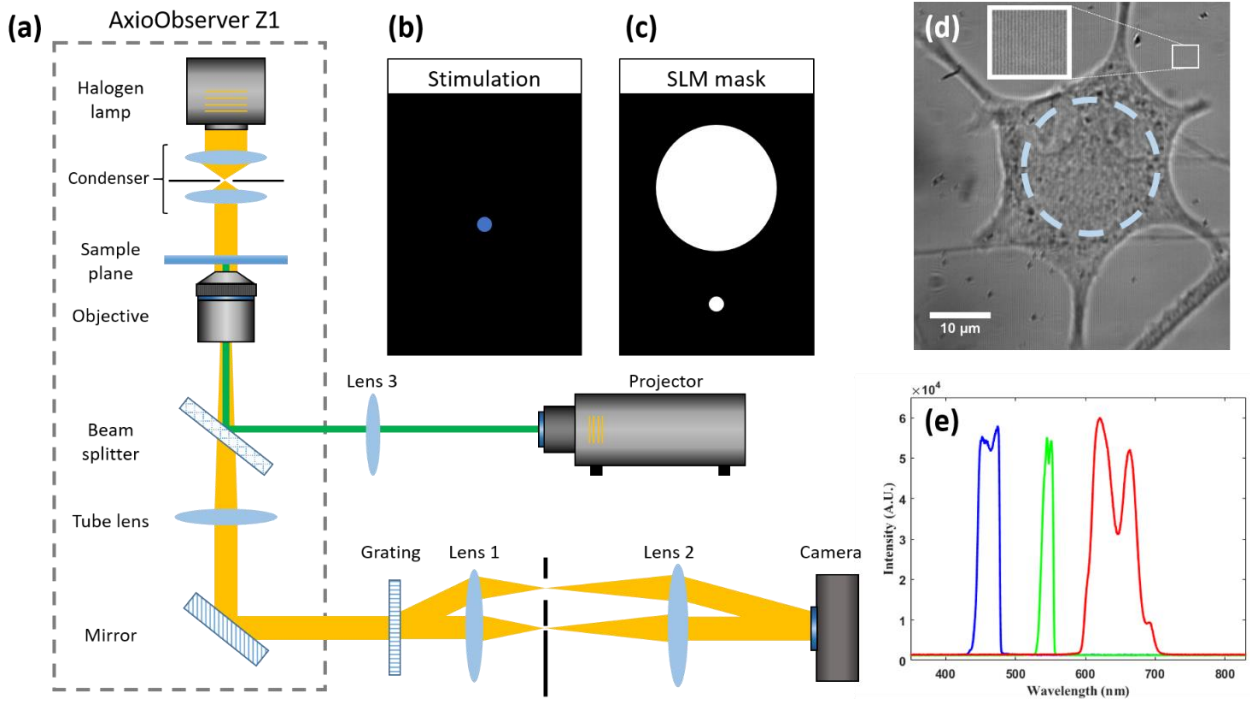

**2.**

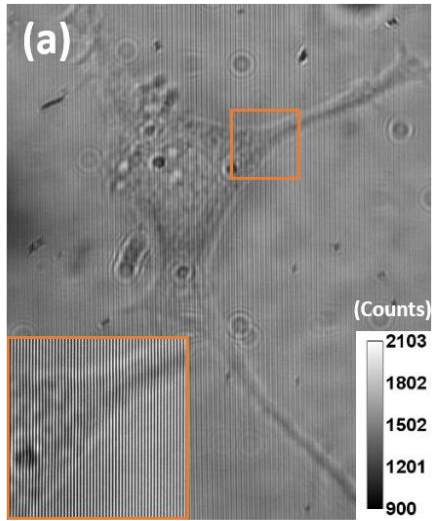 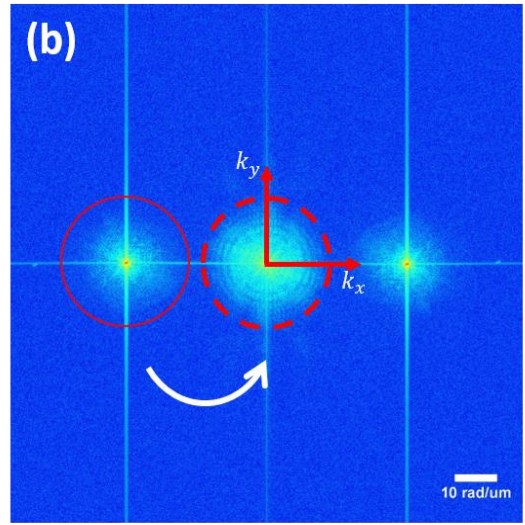

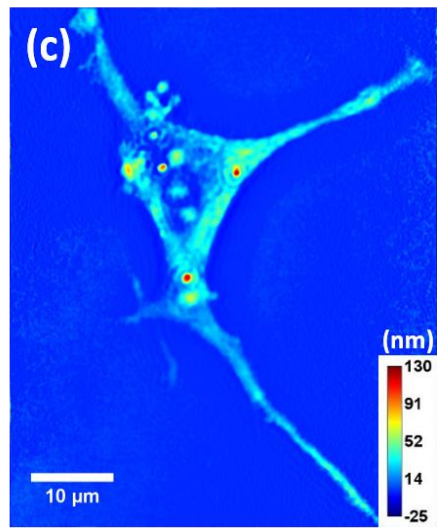 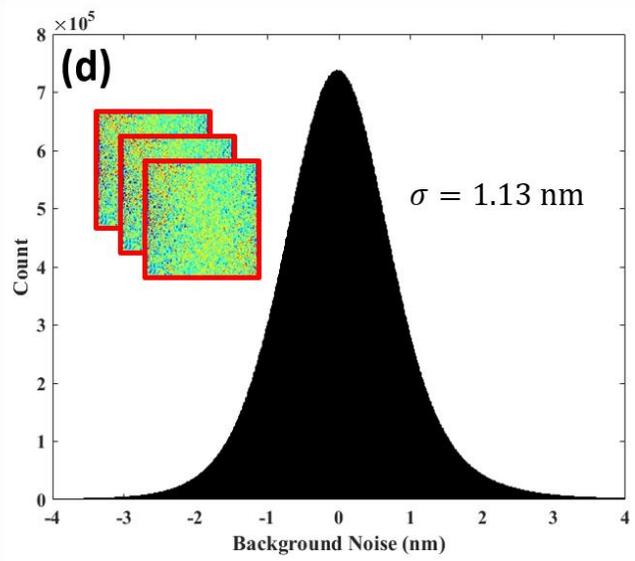

**3.**

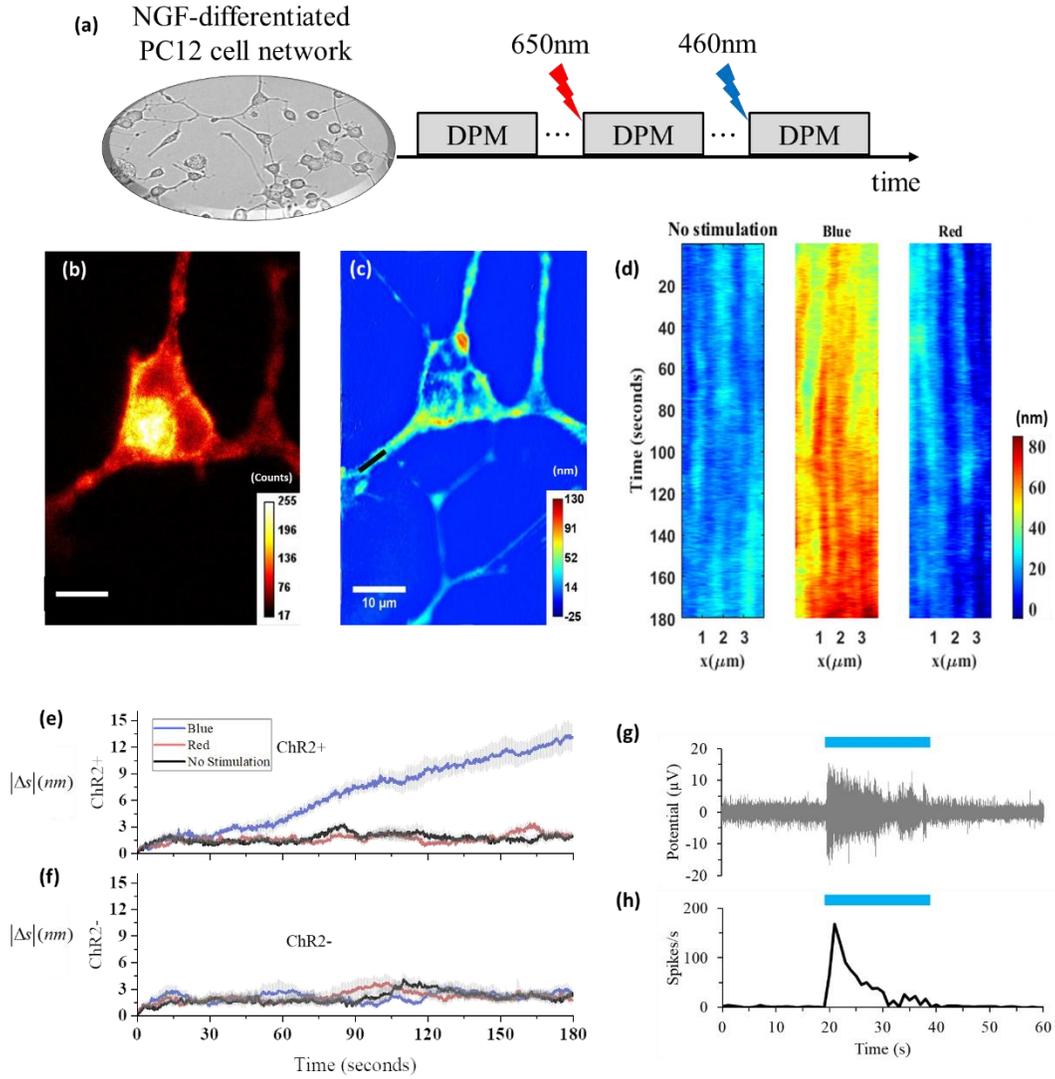

4.

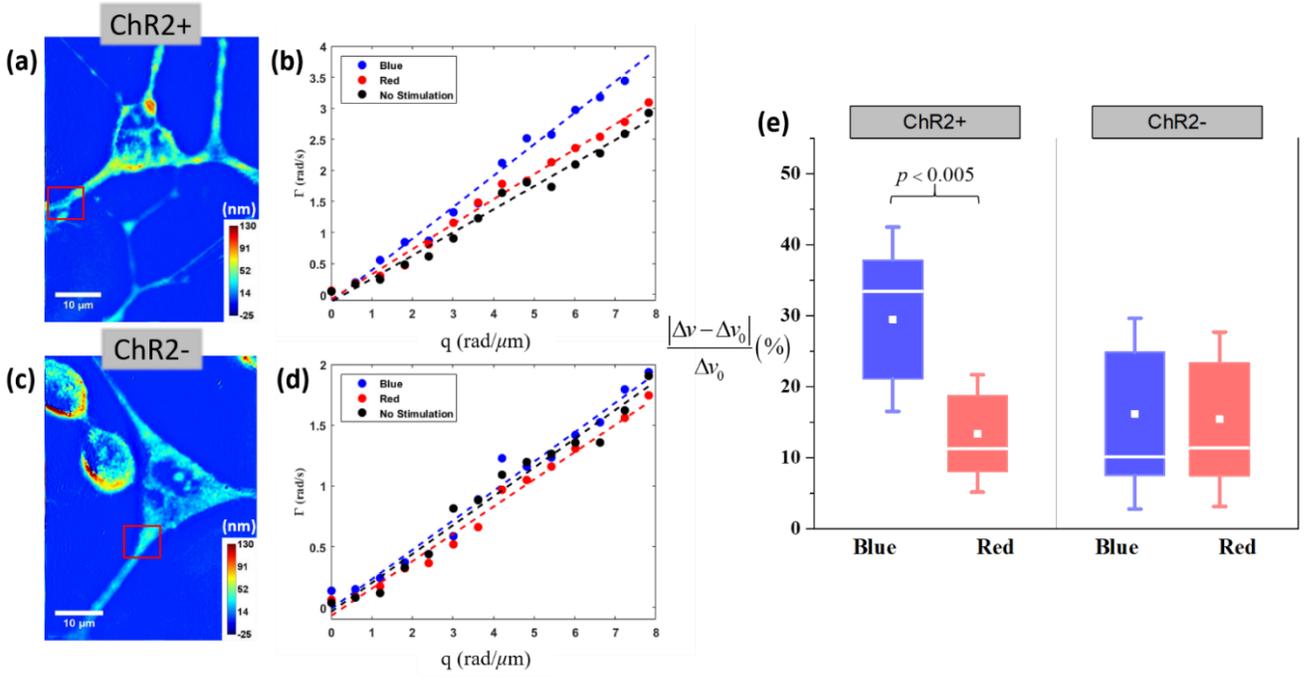

**5.**

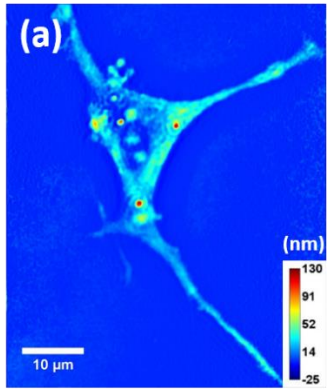
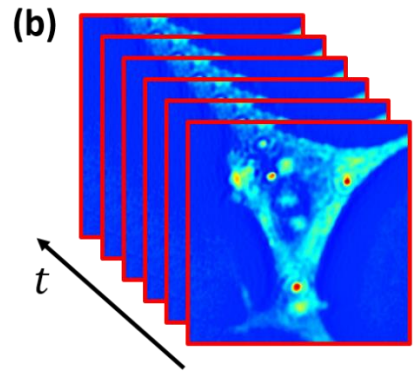
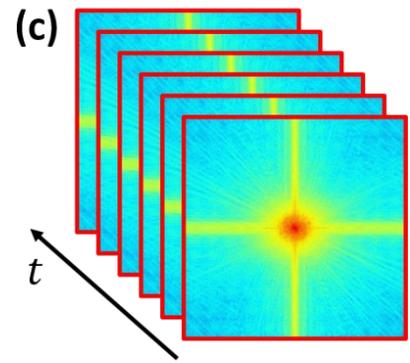
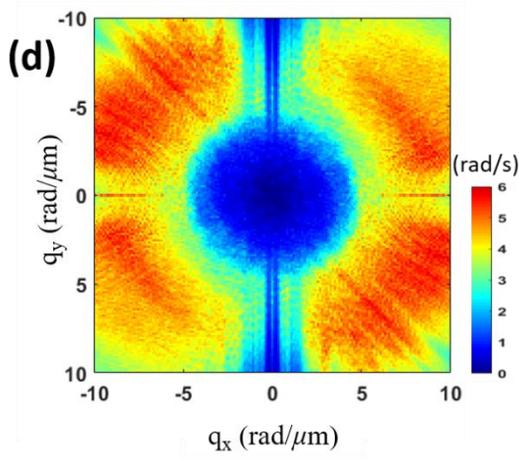
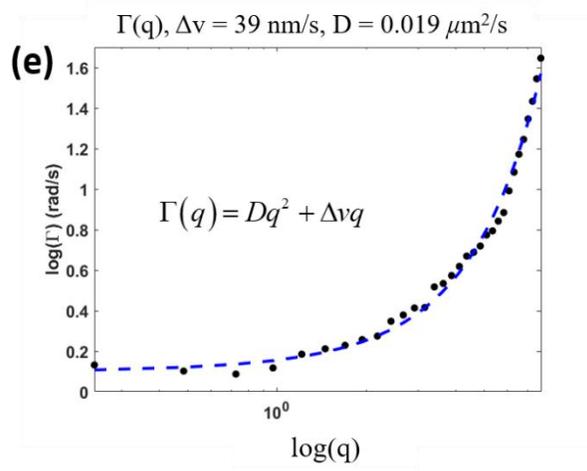

$\Gamma(q) = Dq^2 + \Delta v q$

**Graphical Abstract:**

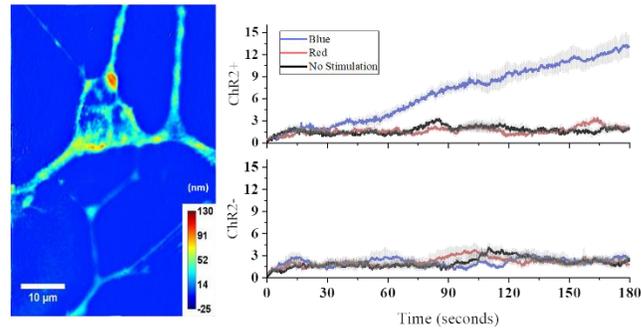

Quantitative phase imaging enables label-free detection of enhanced intracellular activities induced by optogenetic stimulation.